\documentclass[aps,prd,showpacs]{revtex4}

\usepackage{amsfonts}
\usepackage{verbatim}
\usepackage{amsmath}
\usepackage[USenglish]{babel}
\usepackage[dvips]{epsfig}
\usepackage{graphicx}
\usepackage{float}
\usepackage{epsfig,subfig}
\usepackage{epstopdf}
\usepackage{hyperref}
\usepackage{cleveref}
\usepackage{color}
\DeclareMathAlphabet\mathbfcal{OMS}{cmsy}{b}{n}

\begin{document}

\title{Casimir effect between ponderable media as modeled by the standard model extension}

\author{A. Mart\'{i}n-Ruiz}
\email{alberto.martin@nucleares.unam.mx}
\affiliation{Instituto de Ciencias Nucleares, Universidad Nacional Aut{\'o}noma de M{\'e}xico, 04510 M{\'e}xico, Distrito Federal, M{\'e}xico.}

\author{C. A. Escobar}
\email{carlos.escobar@correo.nucleares.unam.mx}
\affiliation{Instituto de Ciencias Nucleares, Universidad Nacional Aut{\'o}noma de M{\'e}xico, 04510 M{\'e}xico, Distrito Federal, M{\'e}xico.}

\begin{abstract}
The CPT-even sector of the standard model extension amounts to extending Maxwell electrodynamics by a gauge invariant term of the form $- \frac{1}{4} (k _{F}) _{\alpha \beta \mu \nu} F ^{\alpha \beta} F ^{\mu \nu}$, where the Lorentz-violating (LV) background tensor $(k _{F}) _{\alpha \beta \mu \nu}$ possesses the symmetries of the Riemann tensor. The electrodynamics in ponderable media is still described by Maxwell equations in matter with modified constitutive relations which depend on the coefficients for Lorentz violation. We study the effects of this theory on the Casimir force between two semi-infinite ponderable media. The Fresnel coefficients characterizing the vacuum-medium interface are derived, and with the help of these, we compute the Casimir energy density. At leading-order in the LV coefficients, the Casimir energy density is numerically evaluated and successfully compared with the standard result. We also found a variety of intriguing effects, such as a non-trivial Kerr effect and the Casimir effect between two phases of the electromagnetic vacuum. We consider a bubble of Lorentz-symmetric (Maxwell) vacuum embedded in the infinite Lorentz-violating vacuum, and we calculate the Casimir energy at leading order, which in this case is quadratic in the LV coefficients. The Casimir force can be positive, zero, or negative, depending on the relative strengths between the LV coefficients.
\end{abstract}

\maketitle

\section{Introduction}

Lorentz symmetry is one of the cornerstones of modern physics, and it stands as one of the basic foundations of general relativity (GR) and the standard model (SM) of particle physics. Both Lorentz covariance to GR and the CPT theorem to the SM involve deeply the Lorentz symmetry. This symmetry has been tested within the energy range currently available, with no violation detected \cite{Cotas}. The question that naturally arises is whether the Lorentz invariance holds exactly or to what extent it holds. Motivated by the unknown underlying theory of quantum gravity \cite{quantum1,quantum2}, the possible Lorentz symmetry violation effects have been investigated from various theories, for example, noncommutative field theories \cite{Noconmu}, loop quantum gravity \cite{QG1}, string theory \cite{String2}, brane-worlds scenarios \cite{Bworld}, condensed matter analogues of ``emergent gravity'' \cite{emergentG}, among others Lorentz violating scenarios \cite{modelos1,modelos2}.

The standard model extension (SME) initiated by Kosteleck\'{y} and Colladay \cite{SME,SME2} is an effective field theory that considers the standard model and general relativity plus all possible Lorentz-violating (LV) coefficients (generated as vacuum expectation values of some basic fields belonging to a more fundamental theory, such as string theory \cite{Strings}) that yield Lorentz scalars (as tensor contractions with the standard fields) under observer Lorentz transformations. One strong motivation to study the SME is because Lorentz violation is a promising candidate signal for Planck-scale physics, and its detection could shed light on the possible route to quantum gravity. Nowadays, Lorentz violation has not been detected in experimental tests, making the SME a current active field of research. The search for novel effects arising from these LV terms, and an improvement of the bounds for the magnitude of these LV coefficients constitute two of the main lines of study of the SME \cite{Parametriz,Studyphot2,testphotonsector}.

Lorentz violation in the photon sector of the SME has been widely investigated in the literature. The study of light propagation (from the cosmic microwave background) offers an opportunity to test LV effects, such as birefringence and polarization properties. Material media have also been used to test Lorentz invariance within the SME framework \cite{BaileyKost, testMaterial}. However, this area remains partially unexplored. In this paper we provide additional contributions to this subject, and we initiate the study of ponderable media as modeled by the photon sector of the minimal SME. More precisely, we analyze materials which besides its dielectric properties, possess additional optical properties (polarization and magnetization) arising from the LV coefficients of the SME. Our primary goal here is to provide additional theoretical predictions regarding the quantum vacuum in this theory, in particular, we concentrate in calculating the Casimir effect between two planar Lorentz-violating ponderable media. It is worth mentioning that the Casimir effect in LV theories has been considered in Refs. \cite{CasimirSME}, however our analysis is quite different since we deal with ponderable media.

The Casimir effect (CE) \cite{Casimir} is one of the most remarkable consequences of the nonzero vacuum energy predicted by quantum field theory which has been confirmed by experiments \cite{Bressi}. The CE can be defined as the stress (force per unit area) on bounding surfaces when a quantum field is confined to a finite volume of space. The boundaries can be  material media, interfaces between two phases of the vacuum, or topologies of space. In any case, the modes of the quantum fields are restricted, giving rise to a measurable force, i.e. the Casimir force. For a review  see, for example,  Refs. \cite{Milton}. The typical example is of the two uncharged conductive plates in a vacuum, placed a few micrometers apart. The experimental accessibility to micrometer-size physics has motivated the theoretical study of the CE in different scenarios, including the gravitational sector \cite{Quach}, condensed matter systems \cite{CE-TI} and chiral metamaterials \cite{CE-MM}. In the problem at hand, the remarkably strong experimental tests with material media can offer new ways to impose (or find) better bounds to the LV coefficients.

In our Casimir system, we consider two Lorentz-violating ponderable media as modeled by the photon sector of the SME. In order to compute the Casimir energy density, we first derive the Fresnel coefficients which characterizes a planar vacuum-medium interface, a result which is absent in the SME literature. With the help of these, next we use the Lifshitz formula to compute the Casimir energy per unit area between two of these materials. The usual Casimir energy density is corrected by a small term which is linear order in the LV coefficients, yielding to an attractive force (as usual) between the bodies. Another solution presented here is the Casimir effect between two phases of the electromagnetic vacuum: a finite Lorentz-symmetric (Maxwell) vacuum embedded in the infinite Lorentz-violating (SME) vacuum.  In this case, the correction to the Casimir energy is second order in the LV coefficients, and we demonstrate that the sign of the slope of the Casimir energy depends upon the relation between the magnitudes of the LV coefficients, thus implying that the Casimir force could be attractive or repulsive.

This paper is outlined as follows. In section \ref{SMEreview} we briefly review the basics of the photon sector of the SME and we establish the model that we are going to deal with. Using the corresponding modified constitutive relations and imposing the boundary conditions at the vacuum-medium interface, in section \ref{FresnelCoefSec} we calculate the Fresnel coefficients. In section \ref{CasimirSec} we use the reflection matrix previously obtained to compute the Casimir energy density IN two different scenarios: i) between two phases of the electromagnetic vacuum and ii) between two semi-infinite ponderable media. A concluding summary of our results comprises the last section \ref{SummConcl}. Here, Lorentz-Heaviside units are assumed ($\hbar =c=1$), the metric signature will be taken as $\left( +,-,-,-\right) $, and we use the conventions, Greek indices $\mu,\nu=0,1,2,3$, Latin indices $i,j=1,2,3$,  and the Levi-Civit\`{a} symbol $\epsilon ^{0123}=+1$.

\section{The photon sector of the SME}
\label{SMEreview}

\subsection{Vacuum electrodynamics}

The Lagrangian of the pure photon sector of the minimal SME is composed by the usual Maxwell Lagrangian plus the additional terms $\frac{1}{2} (k _{AF}) ^{\kappa} \epsilon _{\kappa \lambda \mu \nu} A ^{\lambda} F ^{\mu \nu}$ (sometimes called the Carroll-Field-Jackiw term \cite{CFJ}) and $- \frac{1}{4}(k _{F}) _{\alpha \beta \mu \nu} F ^{\alpha \beta} F ^{\mu \nu}$. The  LV tensor coefficients $(k _{AF}) ^{\kappa}$ and $(k _{F}) _{\alpha \beta \mu \nu}$ are CPT-odd and CPT-even, respectively. Both terms have received much attention in the literature, and experimental constraints exist for them. The CPT-odd term leads to negative contributions to the canonical energy and therefore is a potential source of instability. Thus we set the coefficient to zero, $(k _{AF}) ^{\kappa} = 0$. This is theoretically consistent with radiative corrections in the SME and is well supported experimentally (stringent constraints on $(k _{AF}) ^{\kappa}$ have been set by studying the polarization of radiation from distant radio galaxies \cite{CFJ}). Accordingly, we will consider only the CPT-even tensor in our subsequent analyses. The relevant Lagrangian is \cite{Parametriz}
\begin{equation}
\mathcal{L}= - \frac{1}{4} F _{\mu \nu} F ^{\mu \nu} - \frac{1}{4} (k _{F}) _{\kappa \lambda \mu \nu} F ^{\kappa \lambda} F ^{\mu \nu} - j _{\mu} A ^{\mu}, 
\label{action}
\end{equation}
where $j _{\mu} = ( \rho, - \mathbf{J} )$ is the 4-current source that couples to the electromagnetic 4-potential $A ^{\mu}$, and $F ^{\mu \nu} = \partial ^{\mu} A ^{\nu} - \partial ^{\nu} A ^{\mu}$ is the electromagnetic field strength, which satisfies the homogeneous equations
\begin{equation}
\epsilon _{\mu \nu \alpha \beta} \partial ^{\nu} F ^{\alpha \beta} = 0,
\label{Bianchi}
\end{equation}
ensuring the gauge invariance of the action under the $U(1)$ gauge transformations  $q A _{\mu} \rightarrow q A _{\mu} +\partial _{\mu} \Lambda$. The dimensionless coefficients $(k _{F}) _{\kappa \lambda \mu \nu}$ introduce Lorentz and CPT symmetry breakdown, and they have the symmetries of the Riemann tensor and a vanishing double trace, which imply a total of 19 independent components. The equations of motion arising from the Lagrange density (\ref{action}) are
\begin{equation}
\partial ^{\mu} F _{\mu \nu} + (k _{F}) _{\kappa \lambda \mu \nu} \partial ^{\mu} F ^{\kappa \lambda} = j _{\nu} ,
\label{FieldEqs}
\end{equation}
which extend the usual covariant Maxwell equations to incorporate Lorentz violation. Current conservation can be verified directly by taking the
divergence at both sides of Eq. (\ref{FieldEqs}), i.e. $\partial ^{\mu} \partial ^{\nu} F _{\mu \nu} + (k _{F}) _{\kappa \lambda \mu \nu} \partial ^{\mu} \partial ^{\nu} F ^{\kappa \lambda} = \partial ^{\nu} j _{\nu}$, where the left-hand side vanishes due to the antisymmetry of $F^{\mu\nu}$ and $(k _{F}) _{\kappa \lambda \mu \nu}$ (in the indices $\mu\nu$).

Certain linear combinations of the coefficients $(k _{F}) _{\kappa \lambda \mu \nu}$ for Lorentz violation simplify the analysis of this theory. One useful set can be written in terms of four $3 \times 3$-matrices \cite{Parametriz}, $\kappa_{DE},\kappa _{HB},\kappa _{DB}$ and $\kappa _{HE}$, defined by
\begin{align}
\label{Definicion-k}
\begin{split}
(\kappa _{DE}) ^{jk} &= - 2 (k _{F}) ^{0j0k} , \\ (\kappa _{HB}) ^{jk} &= \frac{1}{2} \epsilon ^{jpq} \epsilon ^{krs} (k _{F}) ^{pqrs} , \\ (\kappa _{DB}) ^{jk} &= - (\kappa _{HE}) ^{kj} = \epsilon ^{kpq}  (k _{F}) ^{0jpq} . 
\end{split}
\end{align}
The $\kappa$ matrices contain the 19 independents components of the tensor $(k _{F}) _{\kappa \lambda \mu \nu}$ as follows: 11 in the matrices $\kappa _{DE}$ and $\kappa _{HB}$, and 8 in the matrices $\kappa _{DB}$ and $\kappa _{HE}$. These definitions imply that the microscopic equations of motion (\ref{FieldEqs}) can be cast in the form of the Maxwell equations for macroscopic anisotropic media \cite{BaileyKost},
\begin{align}
\label{EcuMaxMatter}
\begin{split}
\nabla \cdot \textbf{D} &= \rho \qquad , \qquad \nabla \times \textbf{H} - \frac{\partial \textbf{D}}{\partial t} = \textbf{J} , \\ \nabla \cdot \textbf{B} &= 0 \qquad , \qquad \nabla \times \textbf{E} + \frac{\partial \textbf{B}}{\partial t} = 0 ,
\end{split}
\end{align}
with the modified constitutive relations 
\begin{align}
\label{RelConstitu}
\begin{split}
\textbf{D} &= (1 + \kappa _{DE}) \cdot \textbf{E} + \kappa _{DB} \cdot \textbf{B}, \\[6pt] \textbf{H} &= (1 + \kappa _{HB}) \cdot \textbf{B} + \kappa _{HE} \cdot \textbf{E} . 
\end{split}
\end{align}
As a consequence, many results from conventional electrodynamics in anisotropic media hold for this Lorentz-violating theory. For example, the energy-momentum tensor takes the standard form in terms of $\textbf{E}$, $\textbf{B}$, $\textbf{D}$ and $\textbf{H}$. On the other hand, the general solution to the wave equation in the Lorentz-violating vacuum is similar to that of a plane wave propagating in an anisotropic medium.

Another parametrization of particular relevance for certain experimental considerations employs the four traceless matrices \cite{Parametriz} 
\begin{align}
\label{parametrization-k}
\begin{split}
(\tilde{\kappa}_{e+})^{jk} &= \frac{1}{2}(\kappa_{DE}+\kappa_{HB})^{jk}, \\  (\tilde{\kappa}_{e-})^{jk} &= \frac{1}{2}(\kappa_{DE}-\kappa_{HB})^{jk} - \delta^{jk} \tilde{\kappa}_{\textrm{tr}}, \\ (\tilde{\kappa}_{o+})^{jk} &= \frac{1}{2}(\kappa_{DB}+\kappa_{HE})^{jk} , \\ (\tilde{\kappa}_{o-})^{jk} &=\frac{1}{2}(\kappa_{DB}-\kappa_{HE})^{jk},
\end{split}
\end{align}
where $\tilde{\kappa}_{\textrm{tr}} = \frac{1}{3} \textrm{Tr}(\kappa _{DE})$ is a single coefficient. The matrix $\tilde{\kappa}_{o+}$ is antisymmetric while the remaining matrices are symmetric. The number of independent coefficients in each matrix $\tilde{\kappa}_{o+}$, $\tilde{\kappa}_{o-}$, $\tilde{\kappa}_{e+}$ and  $\tilde{\kappa}_{e-}$ are $3$, $5$, $5$ and $5$, respectively. The remaining independent coefficient is contained in the scalar $\tilde{\kappa}_{\textrm{tr}}$. The above parametrization provides a way to split the coefficients into birefringent and nonbirefringent sectors. The $\tilde{\kappa}_{e+}$ and $\tilde{\kappa}_{o-}$ coefficients which lead to birefringence of light have been strongly bounded (at the level of one part in $10^{32}$ and one part in $10 ^{37}$ respectively) from spectropolarimetry of cosmological distance sources \cite{cotasbirref}. For this reason, throughout this work we disregard the birefringent sector by considering the limit $\tilde{\kappa}_{e+}=\tilde{\kappa}_{o-}=0$. Under this assumption, the constitutive relations can be expressed as
\begin{align}
\label{ConstitSMEusual}
\begin{split}
\textbf{D} &= (1 + \tilde{\kappa}_{\textrm{tr}} + \tilde{\kappa} _{e^{-}}) \cdot \textbf{E} + \tilde{\kappa}_{o^{+}} \cdot \textbf{B}, \\[6pt] \textbf{H} &= (1 - \tilde{\kappa}_{\textrm{tr}} - \tilde{\kappa}_{e^{-}}) \cdot \textbf{B} + \tilde{\kappa}_{o ^{+}} \cdot \textbf{E} ,
\end{split}
\end{align}
which hold in vacuum. An immediate physical consequence of these modified constitutive relations is the transmutation of the electromagnetic fields: an electric charge in vacuum can generate both electric and magnetic fields \cite{BaileyKost}. This magnetoelectric phenomena is also present in real material media, such as topological insulators (TIs) and chiral metamaterials. In the former, the electromagnetic response of TIs is described by standard Maxwell equations in matter with the modified constitutive relations $\textbf{D} = \boldsymbol{\varepsilon} \cdot \textbf{E} + ( \alpha / \pi ) \boldsymbol{\theta} \cdot \textbf{B}$ and $\textbf{H} = \boldsymbol{\mu} ^{-1} \cdot  \textbf{B} - ( \alpha / \pi ) \boldsymbol{\theta} \cdot \textbf{E}$, where $\alpha$ is the fine structure constant and $\boldsymbol{\theta}$ is the topological magnetoelectric polarizability tensor \cite{Martin}. The crystal symmetry of the TI defines the dielectric tensor $\varepsilon _{ij}$ and the magnetic susceptibility tensor $\mu _{ij}$; however, due to time-reversal symmetry, for TIs the magnetoelectric tensor can only have the form $\theta _{ij} = \theta \delta _{ij}$. In the case of chiral metamaterials, the modified constitutive relations have the form $\textbf{D} = \boldsymbol{\varepsilon} \cdot \textbf{E} + i \kappa \textbf{H}$ and $\textbf{B} = \boldsymbol{\mu} \cdot \textbf{H} - i \kappa \textbf{E}$, where $\kappa (\omega)$ is the Condon model for chiral molecular media \cite{CE-MM}. It is worth mentioning that the constitutive relations for these materials are not derivable from Eqs. (\ref{ConstitSMEusual}) because the coefficients $( k _{F} ) _{\alpha \beta \mu \nu}$ for Lorentz violation have the symmetries of the Riemann tensor, and therefore its components can not be the same. 

\subsection{Electrodynamics in ponderable media}

The main goal of this work is to initiate the analysis of this Lorentz-violating electrodynamics in ponderable media. In fact, one can proceeds along the standard lines of electrodynamics in a medium (averaging the microscopic charge and current distributions). The analogue displacement field $\textbf{D}$ and magnetic field $\textbf{H}$ appearing in the vacuum Eqs. (\ref{EcuMaxMatter}) are replaced with the macroscopic displacement field $\textbf{D} _{\textnormal{\tiny{matter}}}$ and macroscopic magnetic field $\textbf{H} _{\textnormal{\tiny{matter}}}$, defined by
\begin{align}
\begin{split}
\textbf{D} _{\textnormal{\tiny{matter}}} &= (1 + \tilde{\kappa}_{\textrm{tr}} + \tilde{\kappa} _{e^{-}}) \cdot \textbf{E} + \tilde{\kappa}_{o^{+}} \cdot \textbf{B} + \mathbfcal{P} , \\[6pt] \textbf{H} _{\textnormal{\tiny{matter}}} &= (1 - \tilde{\kappa}_{\textrm{tr}} - \tilde{\kappa}_{e^{-}}) \cdot \textbf{B} + \tilde{\kappa}_{o ^{+}} \cdot \textbf{E} -\mathbfcal{M} ,
\end{split}
\end{align}
where $\mathbfcal{P}$ and $\mathbfcal{M}$ are the polarization and magnetization, which can be written in terms of averaged molecular electric and magnetic dipole moments. The Maxwell equations in macroscopic media in the presence of Lorentz violation then still take the form (\ref{EcuMaxMatter}).

The full linear response of a ponderable media to applied electric and magnetic fields can be described by the constitutive relations \cite{BaileyKost}
\begin{align}
\label{RelConstMatter}
\begin{split}
\textbf{D} _{\textnormal{\tiny{matter}}} =& ( \boldsymbol{\varepsilon} + \textbf{M} ^{\textnormal{\tiny{vacuum}}} + \textbf{M} ^{\textnormal{\tiny{matter}}} ) \cdot \textbf{E} \\[5pt] & + ( \textbf{N} ^{\textnormal{\tiny{vacuum}}} + \textbf{N} ^{\textnormal{\tiny{matter}}} ) \cdot \textbf{B} , \\[5pt] \textbf{H} _{\textnormal{\tiny{matter}}} =& ( \boldsymbol{\mu} ^{-1} - \textbf{M} ^{\textnormal{\tiny{vacuum}}} - \textbf{M} ^{\textnormal{\tiny{matter}}} ) \cdot \textbf{B} \\[5pt] & + ( \textbf{N} ^{\textnormal{\tiny{vacuum}}} + \textbf{N} ^{\textnormal{\tiny{matter}}} ) \cdot \textbf{E} ,
\end{split}
\end{align}
where we have defined the matrices $M _{ij} = \tilde{\kappa} _{\textrm{tr}} \delta _{ij} + ( \tilde{\kappa} _{e^{-}} ) _{ij}$ and $N _{ij} = ( \tilde{\kappa}_{o^{+}} ) _{ij}$. Here, the permittivity $\varepsilon _{ij}$ and the permeability $\mu _{ij}$ tensors are understood to be those in the absence of Lorentz violation, the matrices $M _{ij} ^{\textnormal{\tiny{vacuum}}}$ and $N _{ij} ^{\textnormal{\tiny{vacuum}}}$ are the contributions due to the LV coefficients in the absence of matter, and the matrices $M _{ij} ^{\textnormal{\tiny{matter}}}$ and $N _{ij} ^{\textnormal{\tiny{matter}}}$ contain pieces of the induced moments $\mathbfcal{P}$ and $\mathbfcal{M}$ that are leading-order in $(k _{F}) _{\mu \nu \alpha \beta}$ and that may be partially or wholly orthogonal to an applied field. The explicit form of the coefficients $(k _{F} ^{\textnormal{\tiny{matter}}}) _{\mu \nu \alpha \beta}$ depends on the macroscopic medium, and they are induced by the atomic-structure modifications from the Lorentz violation. Indeed, unless the matter is isotropic their values can depend on the orientation of the material. Their form can be established in the standard way by applying the averaging process to an appropriate atomic or molecular model based on the SME.

\begin{figure}
\includegraphics[width=7cm, height=4.5cm]{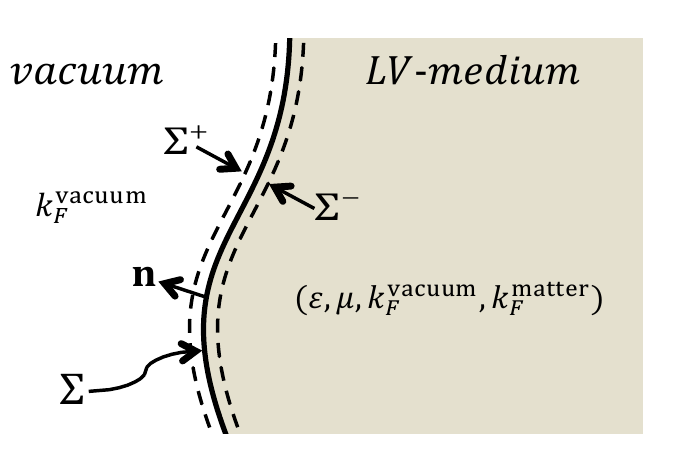}
\caption{\small Geometry of a semi-infinite Lorentz-violating medium in contact with the vacuum.}
\label{Vac-Diel}
\end{figure}

As pointed out by Kosteleck\'{y} \cite{BaileyKost}, in analyzing an experiment, it may be insufficient to replace expressions involving vacuum coefficients with ones involving the sum of vacuum and matter coefficients because the boundary conditions in the presence of matter may induce further modifications. The precise problem we tackle in this paper is the Casimir effect between two Lorentz-violating ponderable media as modeled by the constitutive relations (\ref{RelConstMatter}), in which case the boundaries play a prominent role. To perform the analysis, we require to study the effects of the vacuum-medium interface, which we denote as $\Sigma$, in the propagation of the electromagnetic fields. Assuming that the time-derivatives of the fields are finite in the vicinity of the surface $\Sigma$, together with the absence of free sources on $\Sigma$, the field equations imply the following boundary conditions for the electromagnetic fields at the interface:
\begin{align}
\label{Borderusuales}
\begin{split}
[\textbf{D}] _{\Sigma} \cdot \textbf{n} &= 0 \qquad , \qquad [\textbf{H}] _{\Sigma} \times \textbf{n} = 0, \\[6pt] [\textbf{B}] _{\Sigma} \cdot \textbf{n} &= 0 \qquad , \qquad [\textbf{E}] _{\Sigma} \times \textbf{n} = 0,
\end{split}
\end{align}
which are derived by integrating Eqs. (\ref{EcuMaxMatter}) over a pill-shaped region across $\Sigma$. Here $\textbf{n}$ is the outward unit normal to $\Sigma$ and the notation is $[\textbf{V}] _{\Sigma} = \textbf{V} (\Sigma ^{+}) - \textbf{V} (\Sigma ^{-})$, where $\Sigma ^{+}$ and $\Sigma ^{-}$ are the surfaces just outside and just inside the surface $\Sigma$, as shown in Fig. \ref{Vac-Diel}. If the vacuum matrices $M _{ij} ^{\textnormal{\tiny{vacuum}}}$ and $N _{ij} ^{\textnormal{\tiny{vacuum}}}$ are taken to be zero, then the matter matrices $M _{ij} ^{\textnormal{\tiny{matter}}}$ and $N _{ij} ^{\textnormal{\tiny{matter}}}$ also will be zero, and Eqs. (\ref{Borderusuales}) reduce to the well-known boundary conditions at the interface vacuum-dielectric reported in almost all textbooks on classical electrodynamics. On the other hand, the limits $\varepsilon _{ij} = \delta _{ij}$ and $\mu _{ij} = \delta _{ij}$ (absence of matter) in Eqs. (\ref{RelConstMatter}) imply that these boundary conditions have no effect in the propagation of fields across $\Sigma$, as expected.

To focus the analysis, some simplifying assumptions are adopted in what follows.  For simplicity, we work with isotropic dielectric media with no magnetic properties ($\mu _{ij} = 1$). Thus the dielectric tensor is taken to be of the form $\varepsilon _{ij} (\omega) = \varepsilon (\omega) \delta _{ij}$. Since our primary interest here is the electromagnetic phenomena in ponderable media, let us consider that Lorentz violation is only in the photon sector, and the Lorentz force is conventional. Given that the vacuum LV coefficients have been stringent constrained by experiments, in our model we assume that they can be neglected as compared with the matter LV coefficients, i.e. $( \tilde{\kappa} _{e^{-}} ^{\textnormal{\tiny{matter}}} ) _{ij} \gg ( \tilde{\kappa} _{e^{-}} ^{\textnormal{\tiny{vacuum}}} ) _{ij}$. One simple possible way to support this assumption is by considering that the coupling between the material and the SME background has the simple form $( \tilde{\kappa} _{e^{-}} ^{\textnormal{\tiny{matter}}} ) _{ij} = \varepsilon _{ik} ( \tilde{\kappa} _{e^{-}} ^{\textnormal{\tiny{vacuum}}} ) _{kj}$, such that $\varepsilon _{ij} \gg 1$. In practical terms, the vacuum LV coefficients are set identically zero, and it is appropriate to reduce the model by taking the simplification $\tilde{\kappa}_{o ^{+}} ^{\textnormal{\tiny{vacuum}}} = 0$. This approximation corresponds to consider the complete non-birefringent parity-even sector $(\tilde{\kappa}_{\textrm{tr}},\, \tilde{\kappa} _{e^{-}})$, which encompasses six of the nineteen independent coefficients of $(k _{F}) _{\kappa \lambda \mu \nu}$. This work is not intended to give a complete treatment of material media within the SME framework, which is a meaningful and open problem, but we just give an starting point in the field. Taking these approximations, we may write down the constitutive relations as
\begin{equation}
\textbf{D} = (\boldsymbol{\varepsilon} + \textbf{M}) \cdot \textbf{E} \qquad , \quad\quad\quad \textbf{H} = (\textbf{1} - \textbf{M}) \cdot \textbf{B} ,
\label{constituSMEMatter}
\end{equation}
where
\begin{equation}
M _{ij} = \tilde{\kappa}_{\textrm{tr}} \delta_{ij} + (\kappa _{e ^{-}})_{ij} , \label{defM}
\end{equation}
is a symmetric matrix. In the remainder of the paper, coefficients $\kappa _{e ^{-}}$ (or $M _{ij}$) without label are understood to be those in matter. The relevant vacuum coefficients are explicitly labeled as $\kappa _{e ^{-}} ^{\textnormal{\tiny{vacuum}}}$ (or $M _{ij} ^{\textnormal{\tiny{vacuum}}}$).


\section{Fresnel Coefficients for Lorentz-violating Dielectrics}
\label{FresnelCoefSec}

The Casimir effect is intimately related to the optical properties of two material bodies. There are many ways in which the Casimir effect may be computed. Perhaps the most obvious procedure is to compute the (regularized) zero-point energy in the presence of boundaries. However, a far superior technique is based upon the use of the Green's functions. A particular appealing method is the so-called scattering approach, in which the formula for the force or the interaction energy per unit area can be expressed in terms of the reflection amplitudes at the vacuum-medium interface \cite{Milton}.

In this section we derive the Fresnel coefficients for the case of a planar interface between vacuum and a LV dielectric, a result which to our knowledge is absent in SME literature. With the help of these, in the next section we will analyze the Casimir energy density in two different scenarios: i) between two phases of the electromagnetic vacuum (a finite Lorentz-symmetric (Maxwell) rectangular bubble vacuum embedded in an infinite Lorentz-violating (SME) vacuum) and ii) between two semi-infinite LV dielectric bodies as described by the constitutive relations (\ref{constituSMEMatter}).

The first step is to solve the Maxwell equations in vacuum. Given the rotational symmetry about the $z$-axis, without loss of generality we can choose our coordinate system so that the plane of incidence (defined by $\textbf{k}$ and $\textbf{z}$) coincides with the $xz$ plane. Thus we obtain the required solution by proposing a plane wave solution of the form \cite{optics}
\begin{align}
\label{IncidenteEH}
\begin{split}
& \textbf{E} _{in} = \left( A _{\perp} \textbf{y} + A _{\parallel} \frac{1}{\omega} (k _{z} \textbf{x} - k _{x} \textbf{z}) \right) e ^{i (k _{x} x + k _{z} z - \omega t)}   , \\  & \textbf{H} _{in} = \left( A _{\parallel} \textbf{y} - A _{\perp} \frac{1}{\omega} (k _{z} \textbf{x} - k _{x} \textbf{z}) \right) e ^{i (k _{x} x + k _{z} z - \omega t)} , 
\end{split}
\end{align}
for the incoming wave, and
\begin{align}
\label{reflejadaEH}
\begin{split}
& \textbf{E} _{r} = \left( R _{\perp} \textbf{y} - R _{\parallel} \frac{1}{\omega} (k _{z} \textbf{x} + k _{x} \textbf{z}) \right) e ^{i (k _{x} x - k _{z} z - \omega t)} , \\ & \textbf{H} _{r} = \left( R _{\parallel} \textbf{y} + R _{\perp} \frac{1}{\omega} (k _{z} \textbf{x} + k _{x} \textbf{z}) \right) e ^{i (k _{x} x - k _{z} z - \omega t)} , 
\end{split}
\end{align}
for the reflected wave, provided that $k _{z} = \sqrt{\omega ^{2} - k _{x} ^{2}}$. The Cartesian vectors are represented by $\textbf{x}$, $\textbf{y}$ and $\textbf{z}$. The problem consists in finding the relative amplitudes $A _{\perp}$, $A _{\parallel}$ and $R _{\perp}$, $R _{\parallel}$, which are what define the entries of the reflection matrix.

The second step is to solve Maxwell equations inside the Lorentz violating ponderable medium. The calculation of the plane-wave solutions inside the material is simplified using the coordinate system $(x,y,z)$ attached to the incident wave. If the tensor $M _{ij}$ is defined in a primed coordinate system $(x ^{\prime}, y ^{\prime}, z ^{\prime})$ that forms an angle $\varphi$ with the $x$-axis, the corresponding tensor in the coordinate system $(x,y,z)$ is obtained by rotating it around the $z$-axis by an angle $\varphi$, while the permittivity tensor $\varepsilon _{ij} = \varepsilon \delta _{ij}$ is untouched by the rotation. For simplicity, let us consider that the matrix $M _{ij}$ is defined in the coordinate system $(x,y,z)$. Since there is translational invariance along the interface, $k _{x}$ must be conserved, so transmitted wave can have the form
\begin{align}
\label{EHplanaSME}
\begin{split}
& \textbf{E} _{t} = \textbf{e} \; e ^{i (k _{x} x + q z - \omega t)} , \\ & \textbf{H} _{t} = \textbf{h} \; e ^{i (k _{x} x + q z - \omega t)} , 
\end{split}
\end{align}
where $q$ is the transverse transmitted momentum determined by the dispersion relation of the LV dielectric. Our problem now consists in finding the momentum $q$ and the amplitudes $\textbf{e}$ and $\textbf{h}$ which solve the Maxwell equations. Finally imposing the boundary conditions at the interface we will construct the reflection matrix.

From the Maxwell-Faraday equation $\nabla \times \textbf{E} = - \frac{\partial \textbf{B}}{\partial t}$, with $\textbf{B} = ( \textbf{1} + \textbf{M}) \cdot \textbf{H}$ (which is obtained by inverting the constitutive relation (\ref{constituSMEMatter}) at linear order), one can obtain the conditions for vectors $\textbf{e}$ and $\textbf{h}$, which are
\begin{align}
\label{RotE}
\begin{split}
- q e_{y} &= \omega \sum _{i = x,y,z} ( \textbf{1} + \textbf{M}) _{x i} h _{i} , \\ q e _{x} - k _{x} e _{z} &= \omega \sum _{i = x,y,z} ( \textbf{1} + \textbf{M}) _{y i} h _{i} , \\ k _{x} e _{y} &= \omega \sum _{i = x,y,z} ( \textbf{1} + \textbf{M}) _{z i} h _{i} .
\end{split}
\end{align}
From the Maxwell-Amp\`{e}re law $\nabla \times \textbf{H} = \frac{\partial \textbf{D}}{\partial t}$, with $\textbf{D} = (\boldsymbol{\varepsilon} + \textbf{M}) \cdot \textbf{E} $, one obtains
\begin{align}
\label{RotH}
\begin{split}
q h _{y} = \omega \sum _{i = x,y,z} (\boldsymbol{\varepsilon} + \textbf{M}) _{xi} e _{i} , \\ - ( q h _{x} - k _{x} h _{z} ) = \omega \sum _{i = x,y,z} (\boldsymbol{\varepsilon} + \textbf{M}) _{yi} e _{i} , \\ - k _{x} h _{y} = \omega \sum _{i = x,y,z} (\boldsymbol{\varepsilon} + \textbf{M}) _{zi} e _{i}.
\end{split}
\end{align}
Using the symmetry property $M _{ij} = M _{ji}$, the six previous conditions can be written in the compact matrix form $\textbf{J} \cdot \left( \begin{array}{c} \textbf{e} \\ \textbf{h} \end{array} \right) = 0$, where $\textbf{J}$ is the $6 \times 6$ matrix
\begin{eqnarray}
 \textbf{J} = \left[ \begin{array}{cccccc} \varepsilon + M _{xx} & M _{xy} & M _{xz} & 0 & - \frac{q}{\omega} & 0 \\ M _{xy} & \varepsilon + M _{yy} & M _{yz} & \frac{q}{\omega} & 0 & - \frac{k _{x}}{\omega} \\  M _{xz} & M _{yz} & \varepsilon + M _{zz} & 0 & \frac{k _{x}}{\omega} & 0 \\ 0 & \frac{q}{\omega} & 0 & 1 + M _{xx} & M _{xy} & M _{xz} \\ - \frac{q}{\omega} & 0 & \frac{k _{x}}{\omega} & M _{xy} & 1 + M _{yy} & M _{yz} \\ 0 & - \frac{k _{x}}{\omega} & 0 & M _{xz} & M _{yz} & 1 + M _{zz} \end{array} \right] . \label{matrizfrecuencias}
\end{eqnarray}
The condition for non-trivial solutions, $\det \textbf{J} = 0$, gives us the equation that determines the possible values of $q$ consistent with the Maxwell equations \cite{Itin}. For a given tangential (to the interface) wave vector component $k _{x} \neq 0$, there exist two normal components. At linear order in the LV coefficients the solutions to the characteristic equation are
\begin{align}
\label{frecuencia}
\begin{split}
q ^{\prime} &= \kappa _{z} + \frac{\varepsilon + 1}{\varepsilon} \frac{ \kappa _{z} ^{2} M _{xx} - 2 \kappa _{z} k _{x} M _{xz} + \varepsilon  \omega ^{2} M _{yy} +  k _{x} ^{2} M _{zz}}{4  \kappa _{z}} , \\[8pt] q ^{\prime \prime} &= - \kappa _{z} - \frac{\varepsilon + 1}{\varepsilon} \frac{ \kappa _{z} ^{2} M _{xx} + 2 \kappa _{z} k _{x} M _{xz} + \varepsilon \omega ^{2} M _{yy} + k _{x} ^{2} M _{zz}}{4  \kappa _{z}} ,
\end{split}
\end{align}
where $\kappa _{z} = \sqrt{\varepsilon \omega ^{2} - k _{x} ^{2}}$. Intuitively we expect that the refraction of a wave coming from the vacuum region can only give rise to positive propagating waves, from which we conclude that the solution $q ^{\prime \prime}$ should be discarded. One can further see that the solution $q ^{\prime}$ correctly reduces to the transverse transmitted momentum in an homogeneous dielectric media $\kappa _{z}$ in the limit $M _{ij} \rightarrow 0$.

Now we are ready to impose the boundary conditions. According to Eqs. (\ref{Borderusuales}), tangential components of $\textbf{E}$ and $\textbf{H}$ must be continuous along the interface, thus
\begin{align}
\label{borde1}
\begin{split}
\frac{k _{z}}{\omega} \left( A _{\parallel} - R _{\parallel} \right) &= e _{x} , \\ A _{\perp} + R _{\perp} &= e _{y} , \\ - \frac{k _{z}}{\omega} \left( A _{\perp} - R _{\perp} \right) &= h _{x} , \\ A _{\parallel} + R _{\parallel} &= h _{y} .
\end{split}
\end{align}
One can further check that, using Eqs. (\ref{RotE}) and (\ref{RotH}), the continuity of the normal components of $\textbf{B}$ and $\textbf{D}$ are duplicated by the continuity of $E _{y}$ and $H _{y}$, respectively. Note that the previous system does not depend upon the components $e _{z}$ and $h _{z}$. We can use Eqs. (\ref{RotE}) to obtain a relation where there is no $h _{z}$ dependence, with the result 
 \begin{align}
- \left[ q ^{\prime} (1 + M _{zz}) + k _{x} M _{xz} \right] e _{y} = \omega (1 + M _{xx} + M _{zz}) h _{x} + \omega M _{xy} h _{y} . \label{EQ2}
\end{align}
In a similar fashion, Eqs. (\ref{RotH}) can be combined to eliminate the $e _{z}$ dependence to obtain
\begin{align}
\left[ q ^{\prime} ( \varepsilon + M _{zz}) + k _{x} M _{xz} \right] h _{y} = \omega \varepsilon (\varepsilon + M _{xx} + M _{zz}) e _{x} + \omega \varepsilon M _{xy} e _{y} . \label{EQ1}
\end{align}
The last two equations can be further simplified by using $e_x, e_y, h_x$ and $h_y$ from the Eqs. (\ref{borde1}) to give the matrix equation
\begin{align}
\left[ \begin{array}{cc} \gamma ^{+} (1) & \omega M _{xy} \\  - \omega \varepsilon M _{xy} & \gamma ^{+} ( \varepsilon ) \end{array} \right] \left[ \begin{array}{c} R _{\perp} \\ R _{\parallel} \end{array} \right] = \left[ \begin{array}{cc} \gamma ^{-} (1) & - \omega M _{xy} \\  \omega \varepsilon M _{xy} & \gamma ^{-} ( \varepsilon ) \end{array} \right] \left[ \begin{array}{c} A _{\perp} \\ A _{\parallel} \end{array} \right] \label{AuxMatrix}
\end{align} 
where
\begin{align}
\gamma ^{\pm} ( \varepsilon ) = k _{z} \varepsilon (\varepsilon + M _{xx} + M _{zz} ) \pm q ^{\prime} (\varepsilon + M _{zz}) \pm  k _{x} M _{xz} .
\end{align}
Finally, inverting the $2 \times 2$-matrix of the right hand side of Eq. (\ref{AuxMatrix}) yields the reflection matrix, which can be conveniently written as the sum of two terms,
\begin{align}
\textbf{R} = \textbf{R} ^{(0)} + \textbf{R} ^{(1)} , \label{ReflectionMatrix}
\end{align}
where $\textbf{R} ^{(0)}$ is the zeroth order reflection matrix (in the LV parameters) which is defined by the standard Fresnel coefficients for a planar vacuum-dielectric interface
\begin{align}
\begin{split}
r _{ss} ^{(0)} (\omega , \textbf{k} _{\parallel}) = \frac{k _{z} - \kappa _{z}}{k _{z} + \kappa _{z}} \quad & , \quad r _{pp} ^{(0)} (\omega , \textbf{k} _{\parallel}) = \frac{\varepsilon k _{z} - \kappa _{z}}{\varepsilon k _{z} + \kappa _{z}} , \\[7pt] r _{sp} ^{(0)} (\omega , \textbf{k} _{\parallel}) &= r _{ps} ^{(0)} (\omega , \textbf{k} _{\parallel}) = 0 .
\end{split}
\end{align}
The label $s$ ($p$), equivalent to TE (TM) modes, describes parallel (perpendicular) polarization of the electric field with respect to the plane of incidence. The matrix $\textbf{R} ^{(1)}$ is the linear order correction to the reflection matrix, whose entries are
\begin{align}
\begin{split}
r _{ss} ^{(1)} (\omega , \textbf{k} _{\parallel}) &= \frac{k _{z}}{2 \varepsilon \kappa _{z} (k _{z} + \kappa _{z}) ^{2}}\bigg((3 \varepsilon - 1 ) \kappa _{z} ^{2} M _{xx} - 2 ( \varepsilon - 1 ) \kappa _{z} k _{i} M _{iz} - ( \varepsilon + 1 ) (\varepsilon \omega ^{2} M _{yy} + \textbf{k} ^{2} _{\parallel} M _{zz}) \bigg) , \\ r _{pp} ^{(1)} (\omega , \textbf{k} _{\parallel}) &= \frac{- k _{z}}{2 \kappa _{z} (\varepsilon k _{z} + \kappa _{z}) ^{2}}\bigg(( \varepsilon - 3 ) \kappa _{z} ^{2} M _{xx} - 2 ( \varepsilon - 1 ) \kappa _{z} k _{i} M _{iz} + ( \varepsilon + 1 ) (\varepsilon \omega ^{2} M _{yy} + \textbf{k} ^{2} _{\parallel} M _{zz}) \bigg) , \\  r _{sp} ^{(1)} (\omega , \textbf{k} _{\parallel}) &= - \varepsilon r _{ps} ^{(1)} (\omega , \textbf{k} _{\parallel}) = \frac{2 \omega k _{z} M_{xy}}{(k _{z} + \kappa _{z})(\varepsilon k _{z} + \kappa _{z})} ,
\end{split}
\end{align}
where now $k _{z} = \sqrt{\omega ^{2} - \textbf{k} ^{2} _{\parallel}}$, $\kappa _{z} = \sqrt{\varepsilon \omega ^{2} - \textbf{k} ^{2} _{\parallel}}$, $k _{i} M _{iz} = k _{x} M _{xz} + k _{y} M _{yz}$ and $\textbf{k} ^{2} _{\parallel} = k _{x} ^{2} + k _{y} ^{2}$ as a consequence of the rotational invariance at the vacuum region. In the limit $M _{ij} \rightarrow 0$ the reflection matrix correctly reduces to the reflection matrix for a planar vacuum-dielectric interface, i.e. $\textbf{R} \rightarrow \textbf{R} ^{(0)}$. In such a case there is no polarization mixing since the off-diagonal elements of the reflection matrix $R _{sp} = r _{sp} ^{(1)}$ and $R _{ps} = r _{ps} ^{(1)}$ vanish. However, they are in general not zero and therefore either a $p$- or $s$-polarized incident wave may give rise to both $p$- and $s$-polarized reflected waves. In other words, the polarization plane of the reflected wave is rotated relative to the polarization plane of the incident wave. In particular for the $s$-polarized incident wave, we can get the Kerr angle
\begin{align}
\tan \alpha _{K} ^{(s)} = \frac{r ^{(1)} _{ps}}{r ^{(0)} _{pp} + r ^{(1)} _{pp}} ,
\end{align}
which vanishes in the limit $M _{ij} \rightarrow 0$, as expected. Now we have the main tool to compute the Casimir force between LV dielectric bodies.

We close this subsection by pointing out that our results can be further generalized to the case in which the vacuum LV coefficients are taken into account. Here we argue that these are small as compared with the matter LV coefficients (provided that the dielectric function satisfy $\varepsilon \gg 1$), and thus at first approximation they can be taken to be zero. However, in general we must consider the vacuum anisotropy induced by the LV coefficients $(k _{F} ^{\textnormal{\tiny{vacuum}}}) _{\mu \nu \alpha \beta}$, and a (rigorously derived) molecular or atomic model based on the SME for the matter sector. 


\section{Casimir Effect}
\label{CasimirSec}

There are different techniques for the evaluation of the Casimir force. The original Casimir derivation is based on the calculation of the zero-point energy shift of the electromagnetic field in the presence of the plates. A local formulation based on the evaluation of the vacuum energy-momentum tensor was introduced by Brown and Maclay. A complementary method is the so-called scattering approach, in which the free energy is expressed as a convergent multiple scattering expansion of ray trajectories propagating between the boundaries. Interestingly the formula for the force or the interaction energy per unit area can be expressed in terms of the reflection amplitudes at the vacuum-medium interface. 

\begin{figure}
\includegraphics[width=6cm]{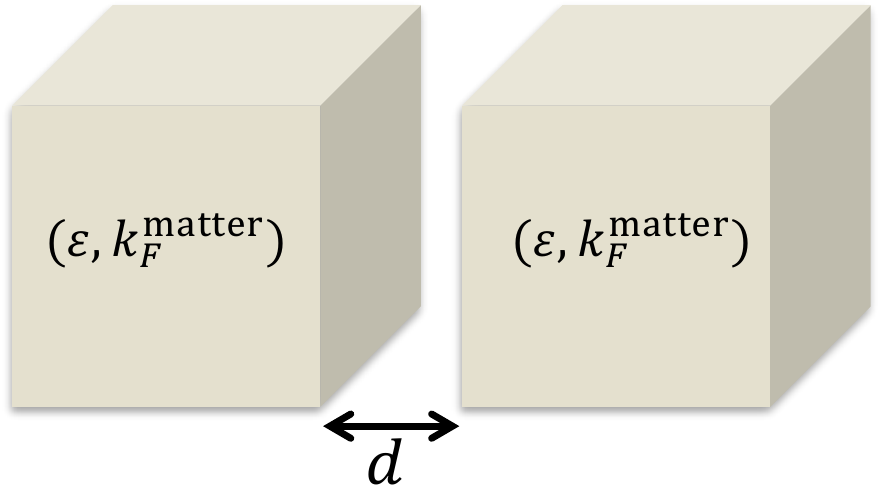}
\caption{\small Lorentz-violating ponderable media separated by a distance $d$.}
\label{LV-DielectricPlates}
\end{figure}

In the previous section we computed the reflection matrix (\ref{ReflectionMatrix}) characterizing the vacuum-LV dielectric interface. With the help of this, in the present section we calculate the Casimir energy density in two different scenarios: i) between two phases of the electromagnetic vacuum and ii) between two semi-infinite Lorentz-violating dielectric bodies. For the situation in which two LV dielectric parallel semi-infinite bodies are placed at a distance $d$ from each other in vacuum, as shown in Fig. \ref{LV-DielectricPlates}, the Casimir energy density per unit area stored in the electromagnetic field between them is given by
\begin{align}
\frac{\mathcal{E} _{C} (d)}{A} = \int _{0} ^{\infty} \frac{d \xi}{2 \pi} \int \frac{d ^{2} \textbf{k} _{\parallel}}{2 \pi} \log \det \left[1 - \textbf{R} _{1} \cdot \textbf{R} _{2} e ^{- 2 k _{3} d} \right] , \label{CasEnergy}
\end{align}
where $A$ is the plate area, $k _{3} = \sqrt{\xi ^{2} + \textbf{k} ^{2} _{\parallel}}$ is the wave vector perpendicular to the plates, $\textbf{k} _{\parallel}$ is the vector parallel to the plates, and $\xi$ is the imaginary frequency defined as $\omega = i \xi$. The matrices $\textbf{R} _{1,2}$ are $2 \times 2$ reflection matrices of media $1$ and $2$ containing the Fresnel coefficients defined as
\begin{align}
\textbf{R} = \left[ \begin{array}{cc} r _{ss} (i \xi , \textbf{k} _{\parallel}) & r _{sp} (i \xi , \textbf{k} _{\parallel}) \\ r _{ps} (i \xi , \textbf{k} _{\parallel}) & r _{pp} (i \xi , \textbf{k} _{\parallel}) \end{array} \right] , \label{RleflectionMatrix}
\end{align}
where $r _{ij}$ describes the reflection amplitude of an incident wave with polarization $i$ which is reflected with polarization $j$. Note that the Fresnel coefficients are evaluated at imaginary frequencies $\omega = i \xi$, and this require the well-known analytic properties of the permittivity in the complex frequency plane. The Casimir force per unit area on the plates is obtained by differentiating expression Eq. (\ref{CasEnergy}) with respect to $d$, i.e. $F = - \partial _{d} \mathcal{E} _{C} (d)$. A positive (negative) force, or equivalently a positive (negative) slope of $\mathcal{E} _{C} (d)$, corresponds to repulsion (attraction) of the plates. Despite the fact that the formula (\ref{CasEnergy}) is commonly used for two homogeneous media, it is still fairly general: it may be applied to dispersive, dissipative, and anisotropic media; all that is needed are the appropriate reflection matrices.


\subsection{Casimir energy between two phases of the electromagnetic vacuum}

Let us consider the Casimir configuration depicted in Fig. \ref{LV-DielectricPlates} in the limiting case in which $\varepsilon = 1$. In this situation, the formula (\ref{CasEnergy}) gives the energy stored in the electromagnetic field between two phases of the electromagnetic vacuum: a finite rectangular bubble of Lorentz-symmetric vacuum embedded in the infinite Lorentz-violating (SME) vacuum. Note that this configuration is possible given that we have assumed that the vacuum LV coefficients are zero, and thus the replacement $(k _{F} ^{\textnormal{\tiny{matter}}}) _{\mu \nu \alpha \beta} \rightarrow (k _{F} ^{\textnormal{\tiny{vacuum}}})$ in Eq. (\ref{ReflectionMatrix}) characterizes the interface between these phases of the vacuum.

The formalism of the previous section can be applied directly to obtain the corresponding reflection matrix by replacing $(k _{F} ^{\textnormal{\tiny{matter}}}) _{\mu \nu \alpha \beta} \rightarrow (k _{F} ^{\textnormal{\tiny{vacuum}}})$. Only in this subsection we understand the coefficients for Lorentz violation to be those in vacuum. We observe that the zeroth order reflection matrix vanishes, i.e. $\textbf{R} ^{(0)} _{\varepsilon = 1} = \textbf{0}$, while the linear order becomes
\begin{align}
\textbf{R} ^{(1)} _{\varepsilon = 1} = \left[ \begin{array}{cc} \mathfrak{r} _{ss} (i \xi , \textbf{k} _{\parallel}) & - \mathfrak{r} _{ps} (i \xi , \textbf{k} _{\parallel}) \\ \mathfrak{r} _{ps} (i \xi , \textbf{k} _{\parallel}) & \phantom{-} \mathfrak{r} _{ss} (i \xi , \textbf{k} _{\parallel}) \end{array} \right] ,
\end{align}
where the Fresnel coefficients are
\begin{align}
\begin{split}
\mathfrak{r} _{ss} (i \xi , \textbf{k} _{\parallel}) &= \lim _{\varepsilon \rightarrow 1} r _{ss} ^{(1)} (i \xi , \textbf{k} _{\parallel}) = \frac{  k _{3} ^{2} M _{xx} - \xi ^{2} M _{yy} + \textbf{k} _{\parallel} ^{2} M _{zz}}{4 k _{3} ^{2}} , \\ \mathfrak{r} _{ps} (i \xi , \textbf{k} _{\parallel}) &= \lim _{\varepsilon \rightarrow 1} r _{ps} ^{(1)} (i \xi , \textbf{k} _{\parallel}) = \frac{\xi  M _{xy}}{2 k _{3}} . 
\end{split}
\end{align}
The Casimir energy for this configuration, at quadratic order in the LV parameters, is
\begin{align}
\frac{\mathcal{E} _{C} (d)}{A} = \int _{0} ^{\infty} \frac{d \xi}{2 \pi} \int \frac{d ^{2} \textbf{k} _{\parallel}}{(2 \pi) ^{2}} 2 \left[ \mathfrak{r} _{ps} ^{2} (i \xi , \textbf{k} _{\parallel}) - \mathfrak{r} _{ss} ^{2} (i \xi , \textbf{k} _{\parallel}) \right] e ^{- 2 k _{3} d}. 
\label{CasEnergyVacuum}
\end{align}
This integral can be evaluated in an analytical fashion. We first write the momentum element as $d ^{2} \textbf{k} _{\parallel} = k _{\parallel} d k _{\parallel} d \vartheta$ and integrate $\vartheta \in [ 0 , 2 \pi]$. Next we replace $\xi$ and $k _{\parallel}$ by the plane polar coordinates $\xi = k _{3} \cos \varphi $, and $k _{\parallel} = k _{3} \sin \varphi$ and finally integrate $\varphi \in [ 0 , \pi / 2]$. This procedure leads to the simple resulting integral $\int _{0} ^{\infty} k _{3} ^{2} e ^{-2 k _{3} d} d k _{3} = 1 / (4 d ^{3})$. The Casimir energy then becomes
\begin{align}
\frac{\mathcal{E} _{C} (d)}{A} &= \frac{1}{ 1920 \pi ^{2} d ^{3}} \left[ 10 M _{xx} \left( M _{yy} - 2 M _{zz} \right) - 15 M _{xx} ^{2} + 20 M _{xy} ^{2} + 4 M _{yy} M _{zz} - 3 M _{yy} ^2 - 8 M _{zz} ^{2} \right] . 
\label{CasEnergyVacuum2}
\end{align}
In most cases the resulting Casimir force between two media separated by a vacuum region is attractive. Recently, there is an increased interest in determining whether there is a combination of media 1 and 2 capable of producing a repulsive force \cite{CE-TI, CE-MM}. In the problem at hand, we observe that the sign of the slope of $\mathcal{E} _{C} (d)$ depends upon the relation between the magnitudes of the components of the tensor $M _{ij}$, and it exhibits the same dependence on the distance ($d ^{-3}$) as that between two parallel conducting plates. For example, if all entries are equal ($M _{ij} = M \delta _{ij}$), the Casimir energy becomes negative, i.e. $\frac{\mathcal{E} _{C} (d)}{A} \sim - \frac{1}{160 \pi ^{2} d ^{3}} M ^{2}$. Note however that this model is not allowed within the SME because $(k _{F}) _{\mu \nu \alpha \beta}$ have the symmetries of the Riemann tensor. Interestingly we can consider a different configuration in which the Casimir energy becomes positive. If we assume that the diagonal entries of $\textbf{M}$ are equal, i.e. $M _{xx} = M _{yy} = M _{zz}$, the Casimir energy (\ref{CasEnergyVacuum2}) reduces to
\begin{align}
\frac{\mathcal{E} _{C} (d)}{A} &= \frac{1}{480 \pi ^{2} d ^{3}} \left( 5 M ^{2} _{xy} - 8 M ^{2} _{xx} \right) , \label{CasEnergyVacuumPart}
\end{align}
which is positive for $\vert M _{xy} \vert > \sqrt{8/5} \vert M _{xx} \vert$, zero for $\vert M _{xy} \vert = \sqrt{8/5} \vert M _{xx} \vert$, and negative for $\vert M _{xy} \vert < \sqrt{8/5} \vert M _{xx} \vert$.

%


\subsection{Casimir energy between two Lorentz-violating ponderable media}

Let us consider the setup depicted in Fig. \ref{LV-DielectricPlates}, in which we have two semi-infinite dielectric bodies as modeled by the photon sector of the SME. According to Eq. (\ref{CasEnergy}), for this case the zero-temperature Casimir energy per unit area at linear order in the LV parameters is
\begin{align}
\frac{\mathcal{E} _{C} (d)}{A} &= \int _{0} ^{\infty} \frac{d \xi}{2 \pi} \int \frac{d ^{2} \textbf{k} _{\parallel}}{( 2 \pi ) ^{2}} \left[ I  ^{(0)} ( \xi , \textbf{k} _{\parallel} ) + I  ^{(1)} ( \xi , \textbf{k} _{\parallel} ) \right] , \label{CasEnergyLV-D}
\end{align}
where the zeroth and first order integrands can be written as
\begin{align}
\begin{split}
 I  ^{(0)} ( \xi , \textbf{k} _{\parallel} ) &= \log \Gamma (\xi , \textbf{k} _{\parallel}) , \\  I  ^{(1)} ( \xi , \textbf{k} _{\parallel} ) &= \frac{2 e ^{-2 k _{3} d}}{\Gamma (\xi , \textbf{k} _{\parallel})} \left\lbrace r _{pp} ^{(0)} ( i \xi , \textbf{k} _{\parallel} ) r _{ss} ^{(0)} ( i \xi , \textbf{k} _{\parallel} )  \left[ r _{ss} ^{(0)} ( i \xi , \textbf{k} _{\parallel} )  r _{pp} ^{(1)} ( i \xi , \textbf{k} _{\parallel} )  + r _{pp} ^{(0)} ( i \xi , \textbf{k} _{\parallel} )  r _{ss} ^{(1)} ( i \xi , \textbf{k} _{\parallel} )  \right] e ^{-2 k _{3} d}  \right. \\ & \hspace{4cm} \left. - \left[ r _{pp} ^{(0)} ( i \xi , \textbf{k} _{\parallel} ) r _{pp} ^{(1)} ( i \xi , \textbf{k} _{\parallel} ) + r _{ss} ^{(0)} ( i \xi , \textbf{k} _{\parallel} ) r _{ss} ^{(1)} ( i \xi , \textbf{k} _{\parallel} ) \right]  \right\rbrace ,
\end{split}
\end{align}
respectively, where
\begin{align}
\Gamma (\xi , \textbf{k} _{\parallel}) = \left[ 1 - r _{ss} ^{(0)2} (i \xi , \textbf{k} _{\parallel}) e ^{-2 k _{3} d} \right] \left[ 1 - r _{pp} ^{(0)2} (i \xi , \textbf{k} _{\parallel}) e ^{-2 k _{3} d} \right] .
\end{align}
Note that the coefficients $r _{sp} ^{(1)} (i \xi , \textbf{k} _{\parallel})$ and $r _{ps} ^{(1)} (i \xi , \textbf{k} _{\parallel})$ are irrelevant in the calculation of the Casimir force at linear order in the LV coefficients. Thus the required Fresnel coefficients evaluated at the imaginary frequency $\omega = i \xi$ are
\begin{align}
r _{ss} ^{(0)} (i \xi , \textbf{k} _{\parallel}) = \frac{k _{3} - K _{3}}{k _{3} + K _{3}} \qquad , \qquad  r _{pp} ^{(0)} (i \xi , \textbf{k} _{\parallel}) = \frac{\varepsilon (i \xi) k _{3} - K _{3}}{\varepsilon (i \xi) k _{3} + K _{3}} ,
\end{align}
\begin{align}
\begin{split}
r _{ss} ^{(1)} (i \xi , \textbf{k} _{\parallel}) &= \frac{k _{3}}{2 \varepsilon (i \xi) K _{3} (k _{3} + K _{3}) ^{2}} \left\lbrace \left[ 3 \varepsilon (i \xi) - 1 \right] K _{3} ^{2} M _{xx} + \left[ 1 + \varepsilon (i \xi) \right] \left[ \textbf{k} ^{2} _{\parallel} M _{zz} - \varepsilon (i \xi) \xi ^{2} M _{yy} \right] \right\rbrace , \\[7pt]  r _{pp} ^{(1)} (i \xi , \textbf{k} _{\parallel}) &= \frac{- k _{3}}{2 K _{3} \left[ \varepsilon (i \xi) k _{3} + K _{3} \right] ^{2}} \left\lbrace \left[ \varepsilon (i \xi) - 3 \right] K _{3} ^{2} M _{xx} - \left[ 1 + \varepsilon (i \xi) \right] \left[ \textbf{k} ^{2} _{\parallel} M _{zz} - \varepsilon (i \xi) \xi ^{2} M _{yy} \right] \right\rbrace ,
\end{split}
\end{align}
where $k _{3} = \sqrt{\xi ^{2} + \textbf{k} _{\parallel} ^{2}}$, $K _{3} = \sqrt{\varepsilon (i \xi) \xi ^{2} + \textbf{k} _{\parallel} ^{2}} $ and $\varepsilon (i \xi)$ is the permittivity function of the dielectric. One recognizes in the first term of Eq. (\ref{CasEnergyLV-D}) the usual Casimir energy density between two semi-infinite dielectric bodies (without SME), while the second represents the correction from the nonbirefringent parity-even sector of the photon Lagrangian.

The most general phenomenological model to describe the optical response of a dielectric is a sum of oscillators to account for particular absorption resonances. When only one oscillator is considered, the dielectric function evaluated at the imaginary frequency $\omega = i \xi $ can be written as
\begin{align}
\varepsilon (i \xi) = 1 + \frac{\omega _{e} ^{2}}{\xi ^{2} + \omega _{R} ^{2} + \gamma _{R} \xi} . \label{Drude}
\end{align}
In this model, $\omega _{R}$ is the resonant frequency of the oscillator while $\omega _{e}$ accounts for the oscillator strength. The damping parameter $\gamma _{R}$ satisfies $\gamma _{R} \ll \omega _{R}$, playing therefore a secondary role on Casimir physics.

\begin{figure}
\begin{center}
\subfloat[\label{CasEnergy1}]{\includegraphics[width = 8.5cm]{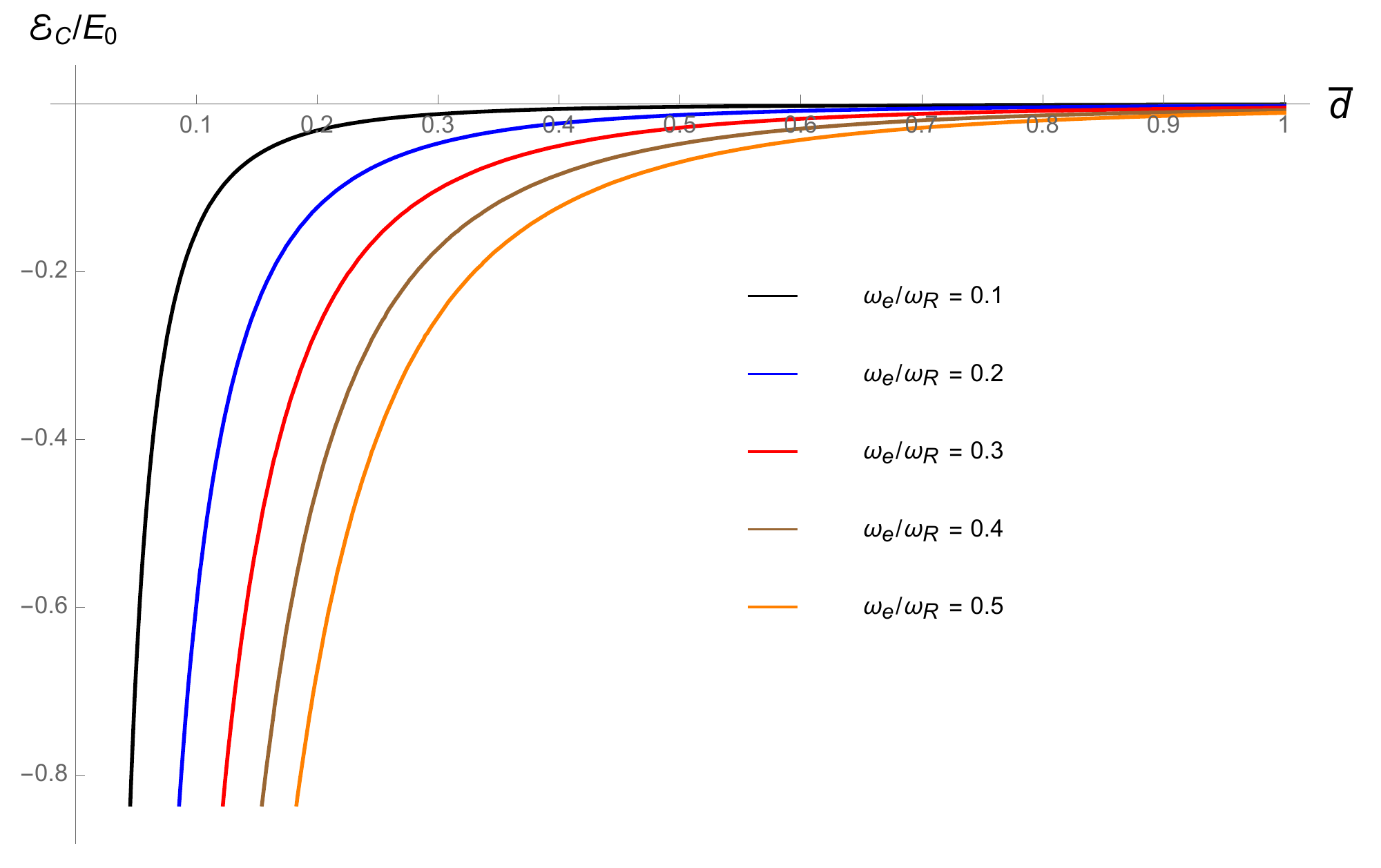}}
\subfloat[\label{CasEnergy2}]{\includegraphics[width = 8.5cm]{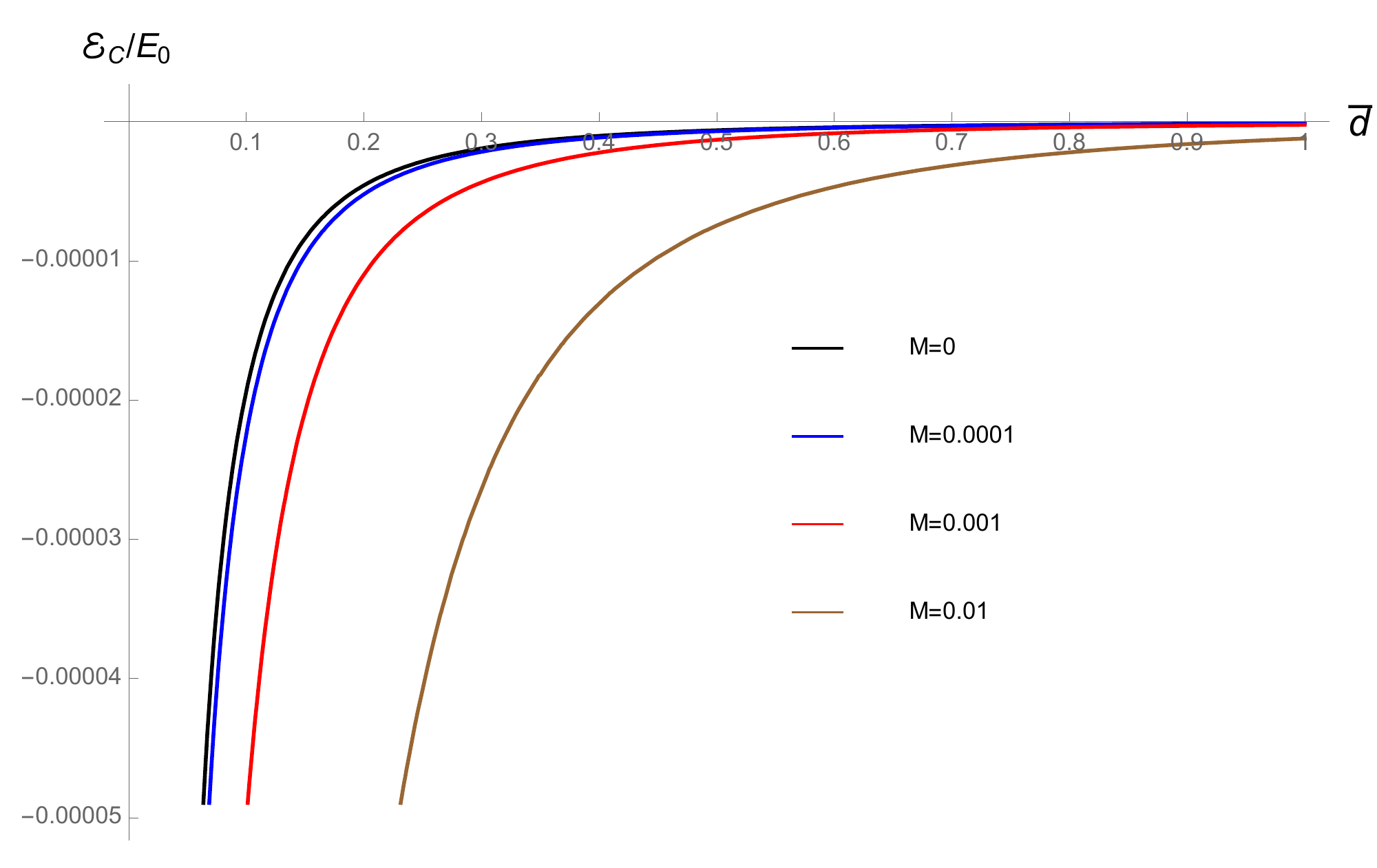}} 
\caption{\small Casimir energy density [in units of $E _{0} = A \omega _{R} ^{3} / (2 \pi) ^{2}$] as a function of the dimensionless distance $\overline{d}$. In (a) $M _{ij} = \delta _{ij}$ is fixed and the different colors correspond to different values of $\omega _{e} / \omega _{R}$.  In (b) $\omega _{e} / \omega _{R} = 0.45$ is fixed and the different colors correspond to different values $M$ (where $M _{ij} = M \delta _{ij}$). }
\end{center}
\label{CasEnergyyyyyyyy}
\end{figure}

One can introduce this model in Eq. (\ref{CasEnergyLV-D}) and study the behavior of the Casimir energy as a function of the two parameters of the model: the LV coefficients $M _{ij}$ and the ratio $\omega _{e} / \omega _{R}$. In this case $\mathcal{E} _{C} (d)$ is a complicated integral expression of the parameters which can not be evaluated analytically, however we can perform a numerical analysis. In this scheme we will investigate two aspects: i) the effect of modifying the relative strength of $\omega _{e} / \omega _{R}$ for $M _{ij} = \delta _{ij}$ (unrealistic model) and ii) the effect of the LV parameters $M _{ij}$ for the fixed value $\omega _{e} / \omega _{R} = 0.45$, which is appropriate for a dielectric with cubic crystal structure.

In what follows, we have rescaled all quantities in units of $\omega _{R}$. Our results are summarized in Figs. \ref{CasEnergy1} and \ref{CasEnergy2} where the Casimir energy [in units of $E _{0} = A \omega _{R} ^{3} / (2 \pi) ^{2}$] is plotted against the dimensionless distance $ \overline{d} = d \omega _{R}$ (recall that we are working in natural units, otherwise $ \overline{d} = d \omega _{R} / c$). \Cref{CasEnergy1} shows the rescaled Casimir energy density as a function of $\overline{d}$ for different values of the ratio $\omega _{e} / \omega _{R}$ and $M _{ij} = \delta _{ij}$. One can further verify that the Casimir energy vanishes for $\omega _{e} / \omega _{R} = 0$. This is so because $r _{ss} ^{(0)} (i \xi , \textbf{k} _{\parallel}) = r _{pp} ^{(0)} (i \xi , \textbf{k} _{\parallel}) = 0$ in this limit, therefore the LV coefficients appears in the Casimir energy at quadratic order, as we discussed in the previous section.  We observe that increasing the ratio $\omega _{e} / \omega _{R}$ shifts the energy toward smaller (negative) values with respect to the standard Casimir energy ($M _{ij} = 0$). However, \cref{CasEnergy1} shows schematically the modification to the Casimir energy for $M _{ij} = \delta _{ij}$, and it is well known that $k _{e ^{-}}$ and $\tilde{\kappa}_{\textrm{tr}}$ are constrained to have very tiny values, thus implying that the distance between the curves is suppressed by these small factors. \Cref{CasEnergy2} shows the Casimir energy density as a function of $\overline{d}$ for different values of $M$ and $\omega _{e} / \omega _{R} = 0.45$, with $M _{ij} = M \delta _{ij}$. We observe that increasing $M$ shifts the energy toward smaller (negative) values with respect to the usual Casimir energy. This plot shows that the energy for $M = 10 ^{- 4}$ is very close to the energy with $M=0$. Nevertheless $M$ has been bounded to be smaller than $10 ^{- 17}$, thus implying that the shift from Casimir energy is undetectable.


\section{Summary and conclusions}
\label{SummConcl}
Although the standard model (SM) is phenomenologically successful, the search for a theory beyond the SM is motivated by the fact that it suffers from some theoretical inconsistencies, and from some long-standing unresolved problems. Searching for Lorentz violation has attracted much attention since it is a promising candidate signal for Planck-scale physics, and its detection could shed light on the possible route to quantum gravity. In this scenario, the SM is viewed as a low-energy limit of this more fundamental theory. The standard model extension (SME) is an effective field theory that incorporates Lorentz-violating (LV) coefficients to the standard model and general relativity, which can be tested by experiments. The SME has been in constant development in recent years and all new alternatives to find better bounds for the LV coefficients are of a great value.

The SME has been the focus of various experimental studies, including ones with photons, electrons, protons, and neutrons. Regarding the photon sector, it has been studied radiative corrections, photon splitting and vacuum \v{C}erenkov and synchrotron radiation. Despite of the experimental efforts, no evidence for Lorentz violation has yet been found. Material media have also been used to test Lorentz invariance, however, this area remains partially unexplored. In this work we aim to fill in this gap. 

We have presented an analysis of the Casimir effect between ponderable media in a Lorentz-violating electrodynamics derived from the photon sector of the SME which includes a CPT-even term of the form $- \frac{1}{4} (k _{F}) _{\alpha \beta \mu \nu} F ^{\alpha \beta} F ^{\mu \nu}$, where the Lorentz-violating background tensor $(k _{F}) _{\alpha \beta \mu \nu}$ possesses the symmetries of the Riemann tensor. In this theory, ponderable media are described by Maxwell equations in matter with modified constitutive relations of the form $\mathbf{D} = \mathbf{D} ( \mathbf{E} , \mathbf{B})$ and $\mathbf{H} = \mathbf{H} ( \mathbf{E} , \mathbf{B} )$. We restrict ourselves to the non-birefringent sector and we assume that the vacuum LV coefficients are small as compared with the matter LV coefficients, in such a way that we set $(k _{F}) _{\alpha \beta \mu \nu} ^{\textnormal{\tiny{vacuum}}}$ identically to zero. Although the last assumption over simplifies the model, It represents the first advances in the field of Casimir physics between ponderable media within the SME.

We derived the Fresnel coefficients which characterizes the vacuum-medium interface, and using the Lifshitz formula, we calculated the leading-order deviation from the classical expression for the Casimir energy density between two parallel dielectric bodies. The deviation of the Casimir force in the Lorentz-violating theory from its standard quantum electrodynamics value is found to be of linear order in the LV coefficients. Using the Drude's model for the dielectric function, we performed a numerical analysis for the Casimir energy density as a function of $\omega _{e} / \omega _{R}$ and the LV coefficients.

We also found a variety of intriguing effects, such as a non-trivial Kerr effect and the Casimir effect between two phases of the electromagnetic vacuum. We considered a finite bubble of Lorentz-symmetric (usual Maxwell) vacuum embedded in the infinite Lorentz-violating (SME) vacuum, and we calculated the Casimir energy at leading-order, which in this case is quadratic in LV coefficients. We found that the Casimir force can be positive, zero, or negative, depending on the relative strengths between the LV coefficients.

Due to the limited precision of the current experimental measurements of the Casimir force, no useful bounds on the LV coefficients can be obtained from our results. The deviation predicted is of theoretical interest, and would only be useful in setting any significant constraints on the LV coefficients only if the precision of the experimental measurements will increase significantly. It is worth mentioning that the Casimir force between two parallel conducting plates has been studied within the SME \cite{CasimirSME}, and the main finding is that the leading-order contribution is quadratic in the LV coefficients, which makes the experimental sensitivity more weaker that the Casimir force between ponderable media. This is a strong motivation to study material media in the SME framework. Finally we comment that our model can be further generalized to more realistic cases, for example, by considering the nonzero background vacuum LV coefficients which are present in the whole space, both inside and outside the material media. Furthermore, a model for the LV coefficients of the matter sector is absent in the SME literature, and clearly they would be of relevance in the calculation of the Casimir effect. We hope our study motives further investigations regarding the matter sector of the SME.

\acknowledgments

We thank Luis Urrutia, Mauro Cambiaso and Marcos Garc\'{i}a for many valuable discussions, comments and suggestions. This work has been partially supported by the project DGAPA-UNAM  \# IN-104815 and the project CONACyT \# 237503.

\end{document}